\documentclass[reprint,superscriptaddress,nofootinbib,amsmath,amssymb,aps,]{revtex4-1}
\usepackage{graphicx}
\usepackage{bm}
\usepackage{xcolor}
\usepackage[normalem]{ulem}

\begin{document}

\title{Tunable interdot coupling in few-electron bilayer graphene double quantum dots}

\author{L. Banszerus}
\email{luca.banszerus@rwth-aachen.de.}
\affiliation{JARA-FIT and 2nd Institute of Physics, RWTH Aachen University, 52074 Aachen, Germany,~EU}
\affiliation{Peter Gr\"unberg Institute  (PGI-9), Forschungszentrum J\"ulich, 52425 J\"ulich,~Germany,~EU}

\author{A.~Rothstein}
\affiliation{JARA-FIT and 2nd Institute of Physics, RWTH Aachen University, 52074 Aachen, Germany,~EU}

\author{E.~Icking}
\author{S. M\"oller}
\affiliation{JARA-FIT and 2nd Institute of Physics, RWTH Aachen University, 52074 Aachen, Germany,~EU}
\affiliation{Peter Gr\"unberg Institute  (PGI-9), Forschungszentrum J\"ulich, 52425 J\"ulich,~Germany,~EU}

\author{K.~Watanabe}
\affiliation{Research Center for Functional Materials,
National Institute for Materials Science, 1-1 Namiki, Tsukuba 305-0044, Japan
}
\author{T.~Taniguchi}
\affiliation{
International Center for Materials Nanoarchitectonics,
National Institute for Materials Science,  1-1 Namiki, Tsukuba 305-0044, Japan
}

\author{C. Stampfer}
\author{C. Volk}
\affiliation{JARA-FIT and 2nd Institute of Physics, RWTH Aachen University, 52074 Aachen, Germany,~EU}
\affiliation{Peter Gr\"unberg Institute  (PGI-9), Forschungszentrum J\"ulich, 52425 J\"ulich,~Germany,~EU}

\date{\today}

\begin{abstract}
We present a highly controllable double quantum dot device based on bilayer graphene. Using a device architecture of interdigitated gate fingers, we can control the interdot tunnel coupling between 1 to 4 GHz and the mutual capacitive coupling between 0.2 and 0.6 meV, independently of the charge occupation of the quantum dots. The charging energy and hence the dot size remains nearly unchanged. The tuning range of the tunnel coupling covers the operating regime of typical silicon and GaAs spin qubit devices.
\end{abstract}

\maketitle

Spin qubits implemented in semiconductor quantum dots (QDs) are attractive candidates for enabling solid state quantum computing~\cite{Loss1998Jan,Yoneda2017Dec,Watson2018Feb}. In particular, singlet-triplet spin qubits, where the logical qubits are encoded in a two-electron spin system in double quantum dots (DQDs) turned out to be very interesting as they allow fast quantum gate operations avoiding fast microwave pulses~\cite{Petta2005Sep,Foletti2009Dec,Barthel2012Jan,Wu2014Aug,Takeda2020Mar}.
For such qubit systems control over the interdot tunnel coupling and hence the exchange interaction between the electrons in the two coupled QDs is essential~\cite{Levy2002Sep,Martins2016Mar,Reed2016Mar,Takeda2020Mar}. Typical tunnel coupling energies are on the order of 1~GHz for silicon-~\cite{Wu2014Aug} and up to 3~GHz for GaAs-based~\cite{Barthel2012Jan} spin qubits allowing fast quantum gate operations.

Bilayer graphene (BLG) is an attractive host material for spin qubits due to its small spin-orbit and hyperfine interaction, as well as the possibility to open a gate voltage controllable band gap~\cite{Trauzettel2007Feb,Oos2007Dec,Zhang2009Jun}.
The development of ultra-clean van der Waals heterostructures where a BLG sheet is encapsulated in hexagonal boron nitride (hBN)~\cite{Wang2013Nov} and a graphite crystal is used as a back gate~\cite{Overweg2018Jan} has lead to a boost in device quality and has enabled the implementation of well-defined QDs~\cite{Eich2018Jul,Kurzmann2019Jul,Banszerus2020Sep,Tong2020Sep,Banszerus2020Oct} and DQDs~\cite{Eich2018Aug,Banszerus2018Aug,Banszerus2020Mar}.
The device architecture used so far to study the single- to few-electron regime in BLG DQDs is based on one gate per QD, where the interdot tunnel barrier is tuned by stray fields of
the dot-defining gates~\cite{Banszerus2020Mar}.
This inhibits the independent control of the interdot tunneling barriers and the charge occupation of the QDs.
The implementation of separate gates, one controlling the dot occupation and one controlling the tunnel coupling, is also possible for BLG QD devices~\cite{Eich2018Aug,Tong2020Sep} and is a well-established technique in different types of GaAs-based QD devices~\cite{Medford2013Jul,Takakura2014Mar,Volk2019Apr,Hsiao2020May}.
In electron and hole QD systems based on SiMOS, Si/SiGe and Ge/SiGe heterostructures, an additional gate layer implementing interdigitated finger gates has been used for that purpose~\cite{Yang2014Nov,Zajac2015Jun,Liles2018Aug,Eenink2019Dec,Lawrie2020Feb}.

Here, we show independent gate control of the tunnel coupling and the mutual capacitive coupling in a few-electron bilayer graphene double quantum dot device while keeping the dot occupations and the dot size, i.e. the charging energy, constant.
In short, we demonstrate the operation of an advanced device architecture with interdigitated gate fingers that allows for a precise modulation of the band edge profile defining the confinement potential and tunneling rates.

\begin{figure*}[]
	\centering
\includegraphics[draft=false,keepaspectratio=true,clip,width=1\linewidth]{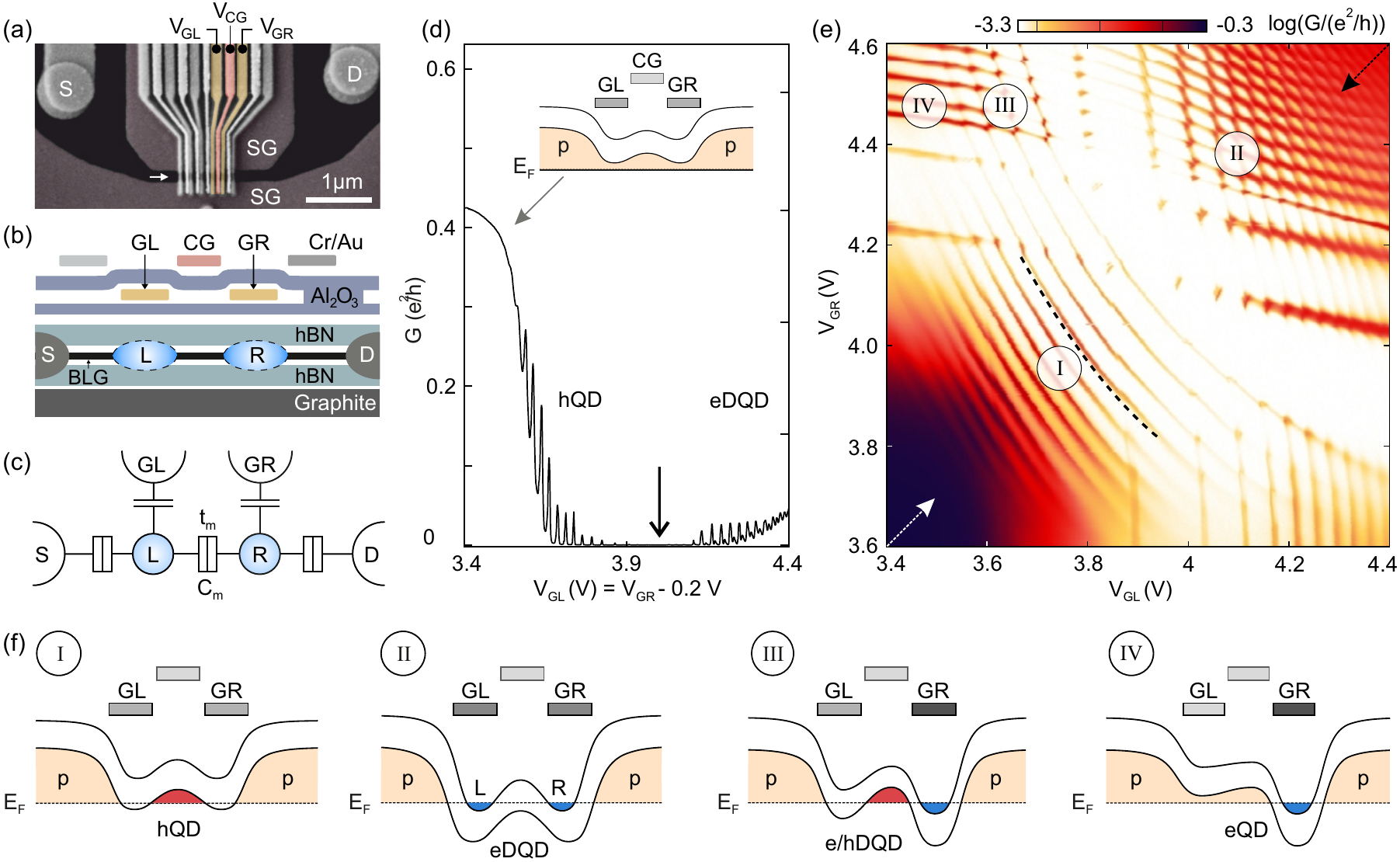}
\caption[Fig01]{
\textbf{(a)} False-color scanning electron micrograph of a fabricated device. The ohmic contacts to the BLG sheet are labelled source (S) and drain (D). The split gates (SGs) define the conducting channel, which can be modulated by voltages applied to the finger gates. The gates used in the following are color coded.
\textbf{(b)} Schematic cross-section through the device along the channel between the split gates (see direction indicated by the arrow in (a)).
\textbf{(c)} Simplified circuit diagram of a DQD device. The QDs are coupled by the mutual capacitance $C_\mathrm{m}$ and the tunnel coupling $t_\mathrm{m}$.
\textbf{(d)} Conductance through the device as a function of the gate voltages $V_\mathrm{GL} = V_\mathrm{GR} - 0.2~$V at $V_\mathrm{CG} = 0$~V, $V_\mathrm{BG} = -1.534~$V, $V_\mathrm{SG} = 1.7~$V and $V_\mathrm{SD} = 500 \, \mathrm{\mu V}$. Inset: Schematics of the band edge profiles in the low gate voltage regime where the Fermi level lies below the valence band edge.
\textbf{(e)}  Charge stability diagram showing the conductance as a function of the gate voltages $V_\mathrm{GL}$ and $V_\mathrm{GR}$. All other voltages as in(d).
\textbf{(f)} Schematics of the valence and conduction band edge profiles along the p-doped channel illustrating the different regimes (I, II, III and IV) set by the left, central and right gate voltages (the darker grey the finger gate, the higher the applied voltage).
}
\label{f1}
\end{figure*}

Fig.~\ref{f1}(a) shows a scanning electron micrograph of a fabricated device.
It consists of a BLG flake, which has been encapsulated between two crystals of hexagonal boron nitride of approximately 25~nm thickness using conventional dry van-der-Waals stacking techniques~\cite{Engels2014Sep,Wang2013Nov}. The heterostructure is placed on a graphite flake, acting as a back gate~\cite{Banszerus2018Aug}.
One-dimensional Cr/Au side contacts are used as ohmic contacts to the BLG~\cite{Wang2013Nov}.
On top of this stack, we deposit metallic split gates (lateral separation of 130~nm) by electron beam lithography, metal evaporation of (5~nm Cr/30~nm Au), and lift-off.
Separated from the split gates by a $25 \, \mathrm{nm}$ thick layer of atomic layer deposited (ALD) $\mathrm{Al_2O_3}$, we fabricate $90 \, \mathrm{nm}$ wide finger gates (FGs) with a pitch of $150 \, \mathrm{nm}$. We use the precursors Trimethylaluminium and H$_2$O in the ALD process and avoid plasma assisted ALD as an O$_2$ plasma can attack the hBN.
To avoid ungated regions along the channel, we fabricate a second layer of FGs, with same width and pitch, separated from the first one by an additional layer of $\mathrm{Al_2O_3}$. A schematic cross section through the heterostructure and the gate stack is shown in Fig.~\ref{f1}(b) where the positions of the left (L) and right (R) QD are highlighted. Fig.~\ref{f1}(c) shows a simplified circuit diagram of a DQD.
All measurements are performed in a ${}^3\mathrm{He}/{}^4\mathrm{He}$ dilution refrigerator at a base temperature of $20 \, \mathrm{mK}$ and an electron temperature of around $60 \, \mathrm{mK}$, using standard DC measurement techniques.
We use a home-built IV-converter with a gain of $10^8$ and a bandwidth of 600~Hz to measure currents in the sub-pA regime. To characterize the BLG, we perform quantum Hall measurements (not shown). We extract a residual doping of $\approx 1.9 \times 10^{10}$~cm$^{-2}$, a charge carrier density inhomogeneiety of $n^*\approx 1.1 \times 10^{10}$~cm$^{-2}$ and a quantum mobility exceeding $20.000$~cm$^2$/(Vs).

In order to form QDs in the extended BLG sheet, we first define a narrow conductive channel by opening a displacement field induced band gap underneath the split gates.
For this, we apply a constant back gate voltage of $V_\mathrm{BG} = -1.534 \, \mathrm{V}$ and a split gate voltage of $V_\mathrm{SG}~=~1.7 \, \mathrm{V}$ resulting in a band gap of around 30~meV in the regions below the SGs and an overall p-doped channel.
Second, we make use of the individual FGs to locally tune the band edges of the gapped BLG with respect to the Fermi level.
For example, when applying a positive voltage on both finger gates GL and GR -- while keeping all other FGs at 0~V -- we tune the band edges such that tunneling barriers form below GL and GR allowing to fully suppress transport through the channel. This is verified by the conductance trace shown in Fig.~\ref{f1}(d) (see black arrow).
For smaller finger gate voltages, we observe regular Coulomb resonances, which we attribute to a hole QD (hQD) below CG, as for large gate voltages we enter the regime of an electron DQD (eDQD).

The different transport regimes become more apparent when investigating the conductance as a function of $V_\mathrm{GL}$ and $V_\mathrm{GR}$ (see Fig.~\ref{f1}(e)); the dashed arrows mark the cross-section shown in Fig.~\ref{f1}(d).
In this charge stability diagram, we highlight the different transport regimes (see labels I, II, III and IV; corresponding schematics of the band edge diagrams are shown in Fig.~\ref{f1}(f)).

\begin{figure*}[]
	\centering
\includegraphics[draft=false,keepaspectratio=true,clip,width=0.85\linewidth]{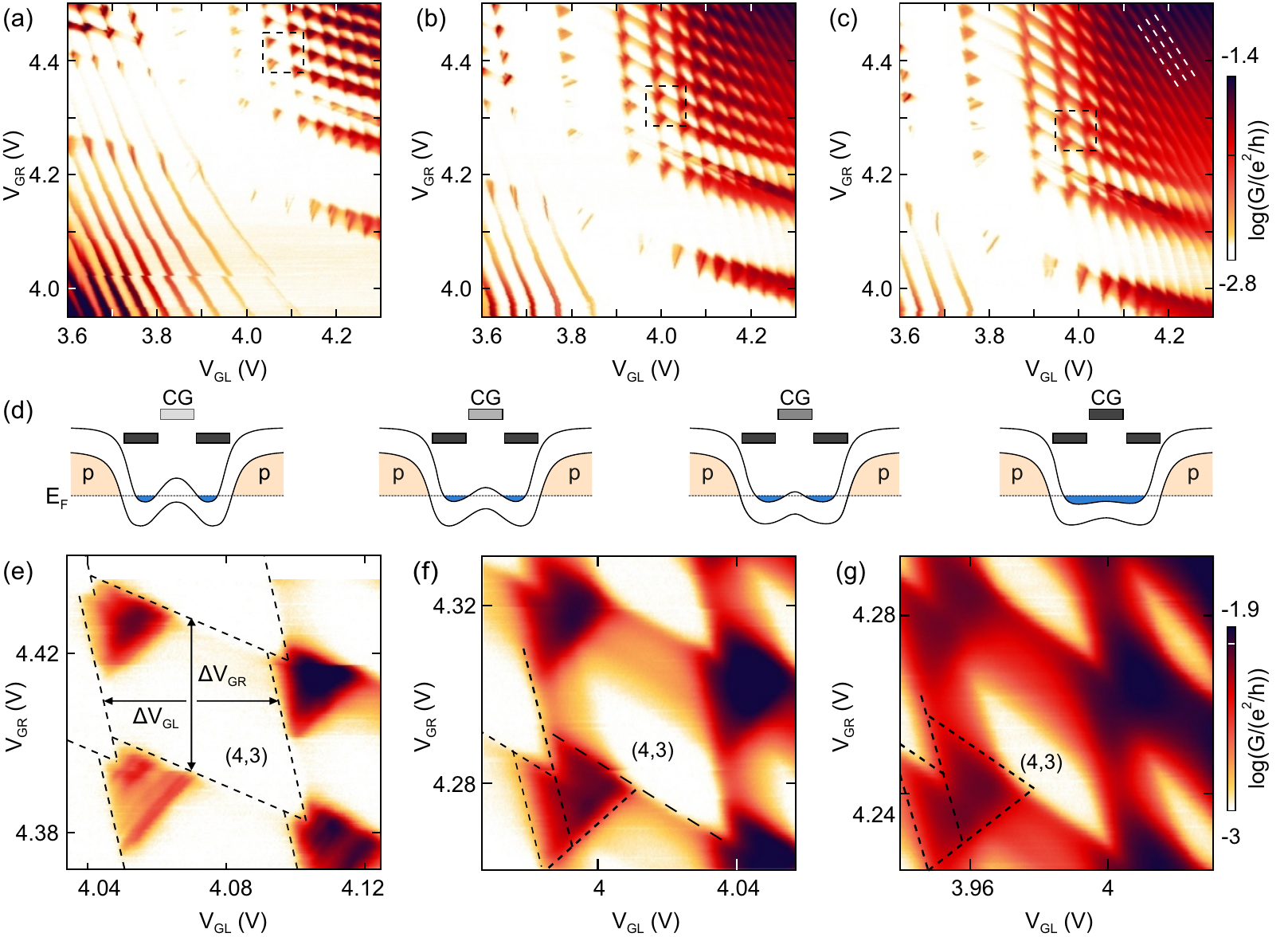}
\caption[Fig02]{
\textbf{(a-c)} Charge stability diagrams showing the conductance as a function of  $V_\mathrm{GL}$ and $V_\mathrm{GR}$ for different $V_\mathrm{CG}=$ 0, 1 and 1.5~V (same range for $V_\mathrm{GL}$, $V_\mathrm{GR}$ in each panel).
\textbf{(d)} Schematics illustrating the effect of $V_\mathrm{CG}$ on the band edge profile in the DQD regime in the panels above.
\textbf{(e-g)}  Stability diagrams showing showing close-ups on the few electron DQD regime as indicated by the dashed boxes in (a)-(c).
Asymmetric split gate voltage of $V_\mathrm{SG,1} = 1.705 \, \mathrm{V}$ and $V_\mathrm{SG,2} = 1.714 \, \mathrm{V}$, a back gate voltage of $V_\mathrm{BG} = -1.547 \, \mathrm{V}$ and a bias voltage of $V_\mathrm{SD}=1$~mV has been applied throughout these measurements.
}
\label{f2}
\end{figure*}

At low voltages around $V_\mathrm{GL} \approx 3.8 \, \mathrm{V}$ and $V_\mathrm{GR} \approx 4 \, \mathrm{V}$ (regime I), we observe hyperbolically shaped charge addition lines indicating the presence of a single hole QD (see dashed line in Fig.~\ref{f1}(e)).
Increasing both gate voltages, the hole QD is depleted more and more due to the capacitive cross-talk of these gates to the QD and we observe the transition to an electron DQD (regime II). The DQD regime shows the characteristic hexagonal pattern of the charge addition lines and extends over a wide voltage range before the increasing tunnel coupling leads to the transition to a single QD.
Interestingly, in an intermediate region, an ambipolar triple QD is formed, where the outer two QDs are occupied by one electron and the inner QD by a single hole.

Furthermore, we can manipulate the band edges to form different ambipolar DQD configurations.
In regime III ($V_\mathrm{GL} \approx 3.6 \, \mathrm{V}$ and $V_\mathrm{GR} \approx 4.5 \, \mathrm{V}$), a hole-electron ambipolar DQD is formed where the horizontal lines indicate charge transitions of the electron QD while the curved lines show transitions of the hole QD. A further reduction of $V_\mathrm{GL}$ at constant $V_\mathrm{GR}$ lifts the tunnel barrier separating the hole QD from the left reservoir leaving only a single electron QD below the right gate (regime IV). The opposite charge configuration can be observed in the bottom right of the shown charge stability diagram.
This measurement proves the versatility of the device which allows smooth transitions between unipolar and ambipolar QD configurations.
Consistent results were also obtained from QD configurations formed by a different pair of finger gates.
The pinch off voltages of the gates in the first finger gate layer (5 out of 6 gates worked) show a spread of $\approx 0.4$~V.

In the following, we focus on the interdot coupling of the electron DQD (regime II).
In order to study the influence of the central gate (CG) located in the second finger gate layer between the gates GL and GR, we measure charge stability diagrams for different $V_\mathrm{CG}$ (see Fig.~\ref{f2}(a)-(c)).
Two significant effects can be observed:
First, all DQD and hole dot transitions in the charge stability diagram are shifted towards lower $V_\mathrm{GL}$ and $V_\mathrm{GR}$ values with increasing $V_\mathrm{CG}$ due to cross capacitances of the CG and the two QDs.
Second, the interdot tunnel coupling in the DQD regime increases as the conduction band edge is pushed more and more towards the Fermi level. This effect is illustrated in the schematics shown in Fig.~\ref{f2}(d).
At high $V_\mathrm{CG}$ (and high $V_\mathrm{GL}$ and $V_\mathrm{GR}$), the tunneling barrier is lifted fully, eventually leading to the formation of a large single QD which manifests in the appearance of diagonal charge addition lines (see e.g. dashed lines in Fig.~\ref{f2}(c)).
Fig.~\ref{f2}(e)-(g) show close-ups of the few electron DQD regime (around the occupation of (4,3) electrons; see dashed rectangles in Fig.~\ref{f2}(a)-(c)). Qualitatively, the effect of increasing interdot tunnel coupling becomes apparent by the broadening of features within the triple points, as well as by the significantly enhanced conductivity along the co-tunneling lines.

\begin{figure}[]
	\centering
\includegraphics[draft=false,keepaspectratio=true,clip,width=0.99\linewidth]{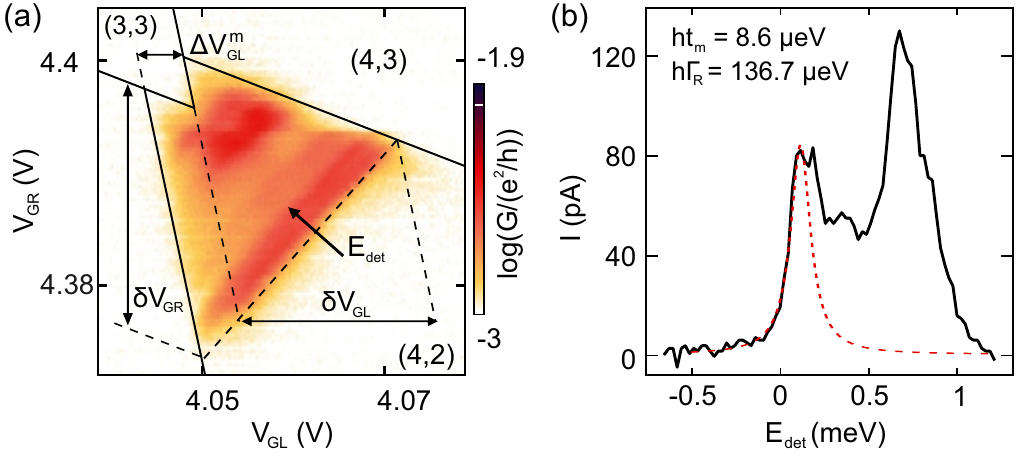}
\caption[Fig03]{
\textbf{(a)}  Charge stability diagram together with a schematic labelling the relevant quantities to extract the lever arm and the capacitive interdot coupling $E^\mathrm{m}$.
\textbf{(b)}  Cut along the detuning axis through the triple point in (a) together with a fit according to eq.~\ref{TunnelCoupling}. $V_\mathrm{SD} = 1 \, \mathrm{mV}$.
}
\label{f3}
\end{figure}

For a quantitative analysis of the impact of the central gate on the interdot coupling and the QD size, we determine the mutual capacitive coupling energy $E^\mathrm{m}$, the interdot tunnel coupling $t_\mathrm{m}$, and the charging energies $E_\mathrm{C}^\mathrm{L,R}$ as a function of $V_\mathrm{CG}$.
Fig.~\ref{f3}(a) shows the charge stability diagram of an individual pair of triple points highlighting the relevant quantities to extract $E^\mathrm{m}$ and $t_\mathrm{m}$.
We determine the charging energy of each of the QDs from the charge stability diagrams as shown in Figs.~\ref{f2}(e)-(g) according to $E_\mathrm{C}^\mathrm{L,R} = \alpha^\mathrm{L,R} \Delta V_\mathrm{GL,GR}$ with the lever arms $\alpha^\mathrm{L,R} = V_\mathrm{SD} / \delta V_\mathrm{GL,GR}$. The mutual capacitive coupling energy is given by $E^\mathrm{m} = \alpha^\mathrm{L} \Delta V^\mathrm{m}_\mathrm{GL}$~\cite{Volk2011Aug}.
The interdot tunnel coupling $t_\mathrm{m}$ can be extracted from current traces recorded along the detuning energy ($E_\mathrm{det}$) axis (see e.g. black arrow in Fig.~3(a)). A representative measurement is shown in Fig.~\ref{f3}(b), where the detuning axis corresponds to a cut  through the triple point shown in Fig.~\ref{f3}(a). Resonances inside the triple point are clearly visible, which correspond to transport involving excited states.
We fit the current through the ground state according to a model assuming a Lorentzian line shape~\cite{Fringes2012Jan,Liu2010May,Wiel2002Dec,Stoof1996Jan} resulting in the limit of $t_\mathrm{m} \ll \Gamma_\mathrm{L,R}$ to
\begin{equation}\label{TunnelCoupling}
I(E_\mathrm{det}) = \frac{4et_\mathrm{m}^2/\Gamma_\mathrm{R}}{1 + (2E_\mathrm{det}/h \Gamma_\mathrm{R})^2},
\end{equation}
where $\Gamma_\mathrm{R,L}$ are the tunnel rates to the left and right lead, respectively.

\begin{figure}[!t]
	\centering
\includegraphics[draft=false,keepaspectratio=true,clip,width=0.99\linewidth]{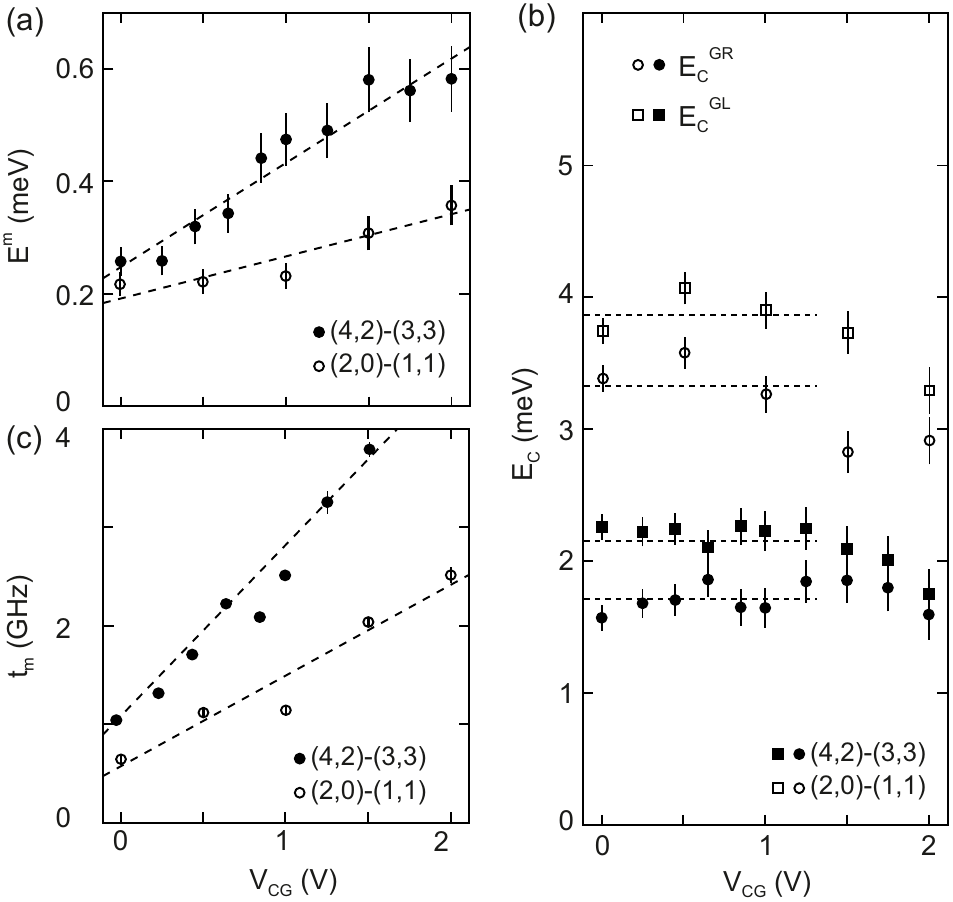}
\caption[Fig04]{
\textbf{(a)}  $E^\mathrm{m}$ measured on the interdot charge transitions (4,2)-(3,3) (filled symbols) and (2,0)-(1,1) (open symbols) as a function of the gate voltage $V_\mathrm{CG}$.
\textbf{(b)}  Charging energy $E_\mathrm{C}$ of the two QDs at the same electron occupations as in (b) as a function of $V_\mathrm{CG}$.
\textbf{(c)}  Interdot tunnel coupling $t_\mathrm{m}$ as a function of $V_\mathrm{CG}$. All dashed lines are guides to the eye.
}
\label{f4}
\end{figure}

The results of the detailed analysis are summarized in Fig.~4.
Fig.~\ref{f4}(a) shows that the mutual capacitive coupling $E^\mathrm{m}$ increases monotonically with $V_\mathrm{CG}$, which can be explained by the
fact that the two electron QDs (L and R) are pushed closer to each other for increasing $V_\mathrm{CG}$ (see sequence shown in Fig.~2(d)). This results in an increase of $C_\mathrm{m}$ and thus to an increase of $E^\mathrm{m}$. Consistently, for lower dot occupations this effect is slightly less pronounced resulting in a lower increase. The observed monotonic behaviour is in contrast to earlier work on a physically etched single-layer graphene DQD~\cite{Molitor2009Jun} and on a gate-defined DQD in an etched graphene nanoribbon~\cite{Liu2010May}, which showed a non-monotonous dependency of $E^\mathrm{m}$ on the gate voltage. Furthermore, in etched BLG DQDs, $E^\mathrm{m}$ increased or decreased with the gate voltage depending an the charge occupation of the DQD~\cite{Volk2011Aug,Fringes2012Jan}.

Fig.~\ref{f4}(b) shows that $V_\mathrm{CG}$ has rather little effect on the charging energy and hence the size of the QDs. From a simplified plate capacitor model approximating the QDs as discs separated from the back gate by 25~nm of hBN, we determine upper limits for the effective QD diameters $d_\mathrm{L} = 220~$nm and $d_\mathrm{R} = 270~$nm in the few electron regime and $d_\mathrm{L} = 174$~nm and $d_\mathrm{R} = 184$~nm in the low electron regime. These estimates are in reasonable agreement with the lithographic dimensions. The pitch of the plunger gates measures 150~nm and the split gates are separated by around 130~nm.

Finally, in Fig.~\ref{f4}(c), we show the tunnel coupling $t_\mathrm{m}$ for a fixed charge carrier occupation for different gate voltages $V_\mathrm{CG}$.
We show that $t_\mathrm{m}$ can be tuned monotonously in the range from 1 to 4~GHz at the (4,2)-(3,3) transition and from 0.7 to 2.5 GHz at the (2,0)-(1,1) transition covering the operating regime for silicon and GaAs spin qubit devices~\cite{Wu2014Aug,Barthel2012Jan}.For larger $V_\mathrm{CG}$ the double dot merges into one large single QD, c.f. higher charge occupation in Fig.~1e).

In conclusion, we studied a BLG QD system where we introduced a second layer of finger gates forming a dense pattern of gates.
We focus on an electron DQD where two dedicated finger gates act as plunger gates controlling the number of charge carriers on each of the QDs from the few-electron regime down to the very last electron. An additional gate, positioned in between those, controls the interdot coupling. Tuning the interdot tunnel coupling in the range of 1 to 4~GHz at a constant charge occupation meets a basic requirement making BLG QD arrays suitable building block for spin qubit devices, which brings BLG closer as a serious quantum technology platform.\\

\textbf{Acknowledgements} The authors thank S.~Trellenkamp, F.~Lentz and D.~Neumaier for their support in device fabrication.
This project has received funding from the European Union's Horizon 2020 research and innovation programme under grant agreement No. 881603 (Graphene Flagship) and from the European Research Council (ERC) under grant agreement No. 820254, the Deutsche Forschungsgemeinschaft (DFG, German Research Foundation) under Germany's Excellence Strategy - Cluster of Excellence Matter and Light for Quantum Computing (ML4Q) EXC 2004/1 - 390534769, through DFG (STA 1146/11-1), and by the Helmholtz Nano Facility~\cite{Albrecht2017May}. Growth of hexagonal boron nitride crystals was supported by the Elemental Strategy Initiative conducted by the MEXT, Japan, Grant Number JPMXP0112101001,  JSPS KAKENHI Grant Numbers JP20H00354 and the CREST(JPMJCR15F3), JST.\\

\textbf{Data availability} The data that support the findings of this study are available from the corresponding author upon reasonable request.

%\bibliography{literature}
%

\clearpage
\end{document}